# Effects of Hard Real-Time Constraints in Implementing the Myopic Scheduling Algorithm

K M. Sakib, M S. Hasan [1], and M A. Hossain [2]

Institute of Information Technology,
University of Dhaka,
Dhaka 1000
muheymin@yahoo.com,

[1] Department of Computer Science & Engineering,
University of Dhaka,
Dhaka 1000,
shahidul_hasan@yahoo.com,

[2] Department of Computing University of Bradford
UK
m.a.hossain1@bradford.ac.uk

*Abstract-*

**Myopic is a hard real-time process scheduling algorithm that selects a suitable process based on a heuristic function from a subset (Window) of all ready processes instead of choosing from all available processes, like original heuristic scheduling algorithm. Performance of the algorithm significantly depends on the chosen heuristic function that assigns weight to different parameters like deadline, earliest starting time, processing time etc. and the size of the Window since it considers only $k$ processes from $n$ processes (where, $k \leq n$). This research evaluates the performance of the Myopic algorithm for different parameters to demonstrate the merits and constraints of the algorithm. A comparative performance of the impact of window size in implementing the Myopic algorithm is presented and discussed through a set of experiments.**

*Keywords-***Original scheduling algorithm, Myopic algorithm, window size, earliest starting time, processing time, deadline of a process, dispatch queue, heuristic function.**

## I. INTRODUCTION

This paper presents the constraints and impact of window size of Myopic scheduling algorithm for hard real-time system. Hard real-time systems are used in time critical applications like avionics, nuclear weapon control, robotics etc [1]. These systems must guarantee that all the processes are completed by their explicit deadlines [2]. If a process cannot meet its deadline, it is discarded. The target of these systems is to determine the best order of the processes so that all or most of the processes meet their deadlines. This requires an efficient scheduling algorithm, which can be performed in two ways – statically and dynamically [3]. In static algorithms, the order of processes and the time they start execution can be determined in advance. Static algorithms are suitable for periodic tasks with hard deadlines [4], [5]. Dynamic algorithms deal with aperiodic processes, whose characteristics are not known a priori [5], [6]. When new tasks arrive, the scheduler selects the most suitable task without any knowledge of the previously scheduled tasks. Once a task has been selected, it is sent to the dispatch queue of the processor. A process is considered to be feasible if the scheduling satisfies its timing constraints and the schedule in which every task is feasible is called feasible schedule [7].

---

This investigation explores the original heuristic scheduling algorithm followed by Myopic scheduling algorithm. The impact of the window size, number of processes and processing time of the Myopic algorithm is explored through a set of experiments. A comparative performance of the impact of window size of the algorithm is evaluated to demonstrate the merits of different window sizes for hard real-time implementation.

This paper is organized as follows. A task model is given in Section II, upon which algorithm performance is measured. Section II also illustrated the basic heuristic based scheduling algorithm and the Myopic algorithm. Section III described the simulation results for varying processes (100 to 1000 processes) and for varying processing time (5-6 to 20-21 time units). Finally, the paper is concluded at Section IV.

## II. ALGORITHMS

### A. The Task Model

Processes are considered to have the following properties [2] to evaluate the constraints and impacts:

- $T_G$ (Generation time): An absolute time when the process is generated or submitted.
- $T_P$ (Processing time): The worst case processing time from starting time of execution.
- $T_D$ (Deadline): An absolute time by which it must complete its execution.
- $\{T_{REQ}\}$ (Resource requirement vector): Resources can be requested exclusively or in shared mode.
- $T_{EST}$ (Earliest start time): An absolute time when a process can begin execution. It must meet the condition $(T_D - T_P) \geq T_{EST} \geq T_G \geq 0$.
- Tasks are aperiodic and non-preemptive.

### B. The Original Heuristic Scheduling

The original heuristic scheduling algorithm starts with an empty schedule and adds processes one by one [2]. It chooses a process from the set of processes $T$, based on a heuristic function, which can consider any of the following formulas:

- $H(T) = Min(T_D)$: The earliest deadline of all processes.
- $H(T) = Min(T_P)$: The process with the shortest processing time.
- $H(T) = Min(T_{EST})$: The earliest $T_{EST}$ carrying process.
- $H(T) = Min(T_D - T_{EST} - T_P)$: The process with the shortest laxity time.
- $H(T) = T_D + W \times T_P$
- $H(T) = T_D + W \times T_{EST}$

Here, $W$ is a weight parameter that controls relative importance between $T_D$ and $T_P$ or $T_{EST}$. The smaller $T_D$ and $T_P$ (or $T_{EST}$) a process has, the more priority it receives. So, the algorithm picks the process with the smallest $H$ (heuristic) value to form the partial schedule. After choosing the first process the schedule becomes a partial schedule. Then the algorithm checks for the strong feasibility. Strong feasibility constraint is satisfied if all the processes in the partial schedule meet their deadlines or timing constraints [1], [2]. If it is not met then the algorithm can take any one of the following steps:

- The algorithm may abort
- It can backtrack and change the latest chosen process etc.

At each step the algorithm includes one process in the partial schedule. For $n$ tasks set there will be $n$ steps and in each step the algorithm will compute $H$ for at best $n$ processes. So, the complexity of the algorithm is $O(n^2)$.

*C. Myopic Scheduling Algorithm*

The Myopic algorithm is explained with the following terms [2]:

- $\{Task\_remaining\}$: the tasks that have not been scheduled
- $N_R$: the number of tasks in the set $\{Task\_remaining\}$
- $K$: Feasibility check window, the maximum number of tasks in $\{Task\_remaining\}$ that will be considered
- $N_K$: Actual number of tasks that are considered, $N_K = Min(k, N_R)$.
- $\{Task\_considered\}$: the first $K$ processes in the $\{Task\_remaining\}$ that are considered.

The tasks in the $\{Task\_remaining\}$ are always kept sorted by increasing order of deadlines, $T_D$. Myopic algorithm works like the Original Heuristic Algorithm with the exception that it applies the heuristic and strong feasibility to only $K$ (where $K \leq n$) processes instead of $n$ processes. This algorithm is called Myopic because it is a shortsighted approach for decision-making [2].

There are $n$ steps for including $n$ processes and in each step it applies heuristic, strong feasibility to only $K$ processes. So, the complexity becomes $O(Kn)$ [1], where $K \leq n$. For small value of K (Window size) the scheduling operation executes faster. But since the algorithm considers only few processes to choose the best one, it exhibits worse performance. On the other hand, if it considers all the processes $K = n$ then the scheduling operation executes slower and becomes the Original scheduling algorithm [2].

### III. SIMULATION DETAILS AND RESULTS

The Myopic algorithm is implemented on a high performance Pentium PC using C++ programming language, assuming the model as discrete time model. For a specific conditions or value of parameters, five observations were taken and presented. These conditions and parameters are described below:

*A. Choice of Heuristic Function*

Among the heuristic functions listed in section II B, $H(T) = T_D + W \times T_{EST}$ considers all the required parameters: deadline, generation time, resource requirement etc. Therefore, this implementation used the heuristic function $H(T) = T_D + W \times T_{EST}$ to judge the performance of the Myopic algorithm for various window sizes $K$ [3] and different values of $W$ under different load.

*B. Effect of $W$*

This parameter controls the relative weight of $T_D$ and $T_{EST}$. If $W = 0$ then the heuristic becomes completely $H(T) = Min(T_D)$ or in other words, earliest deadline first (EDF) algorithm [8]. If, $W$ is set to 1, then $T_D$ and $T_{EST}$ both have the same importance. For the purpose of simplicity, this investigation considered the cases $W = 0.5$ and $W = 1.0$.

## C. Process Properties

Processes in implementing Myopic algorithm are considered as Non-preemptive. Therefore, once a process enters the CPU for execution, no other process can preempt it. Processes are scheduled when the current process is completed. All the processes have random $T_D$, $T_P$, $T_G$ and $T_{REQ}$ values. Processing time is varied from 5 to 21 time units.

## D. Window size range

Window sizes in implementing the algorithm are considered for 2, 4, 6, 8, 10. Window size 1 is not considered, because the Heuristic $H(T) = T_D + W \times T_{EST}$ with window size 1 becomes completely EDF algorithm.

## E. Effect of Load

This section presents the impact of Window size due to variation of number of processes. To demonstrate the impact, loads of 200, 500 and 1000 processes are used for the parameter $W = 0.5$ and $W = 1.0$. The result obtained through a set of experiments is described below.

*Case 1: For 200 processes with processing time 10 to 11 time unit and laxity 100 time unit*

Figure 1(a) and 1(b) show the performance in implementing the Myopic algorithm for 200 processes with processing time 10 to 11 time unit and the value for $W$ is 0.5 and 1.0 respectively.

It is noted from the Figure 1(a) that the number of average completed processes increases slightly for larger window size. It may be due to the fact that for larger window the algorithm is capable of selecting more suitable process from the available processes. For window size 2, the choice for the process is very limited and most probably, for this reason, reading 2, 3, 4 and 5 show the same number of completed processes. A significant level of oscillation is also noted for window size 6, which could be due to the random nature of the processes. From Figure 1(b), it is perceived that the number of completed processes is lower than the performance shown in Figure 1(a). In general, it is reflected from Figure 1(a) & 1(b) that the number of completed processes drops due to the impact of the higher priority of the processing time. It is also noted from Figure 1(b) that instead of oscillation the performance becomes constant at window size 6 and higher.

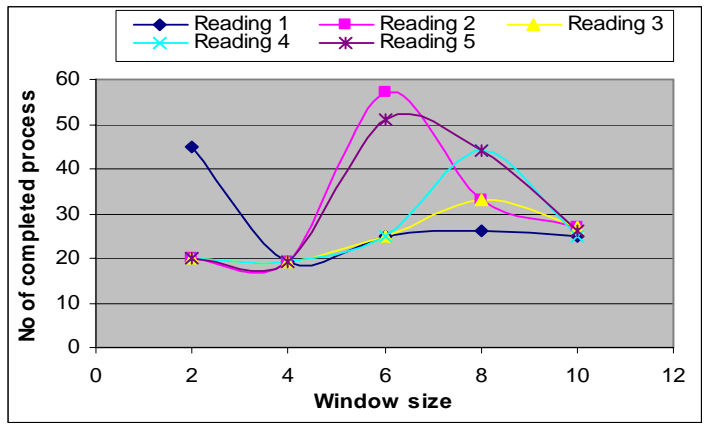

Figure. 1(a). Performance in implementing Myopic algorithm for W=0.5.

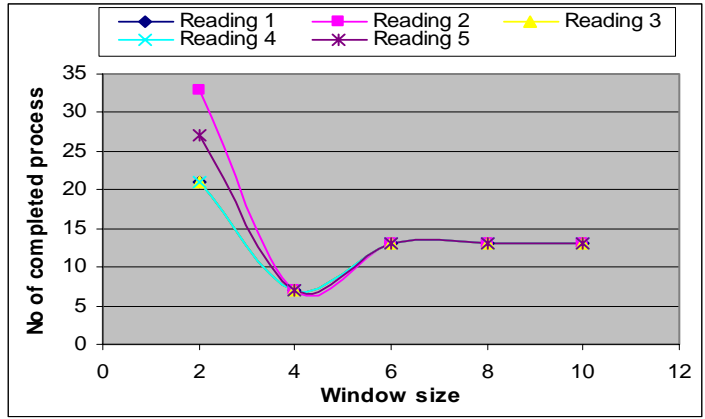

Figure. 1(b).  Performance in implementing Myopic algorithm for W=1.0.

*Case 2: For 500 processes with processing time 10 to 11 time unit and Laxity 100 time unit*

Figure 2(a) and 2(b) depict the performance in implementing the Myopic algorithm for 500 processes with processing time 10 to 11 time unit and the value for $W$ is 0.5 and 1.0 respectively.

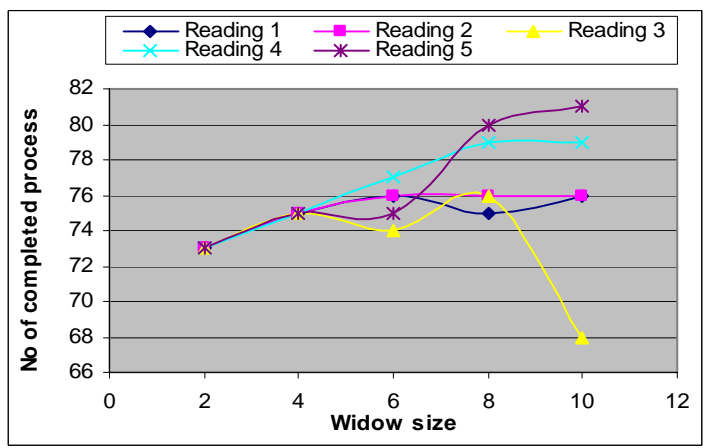

Figure.  2(a).  Performance in implementing Myopic algorithm for W=0.5.

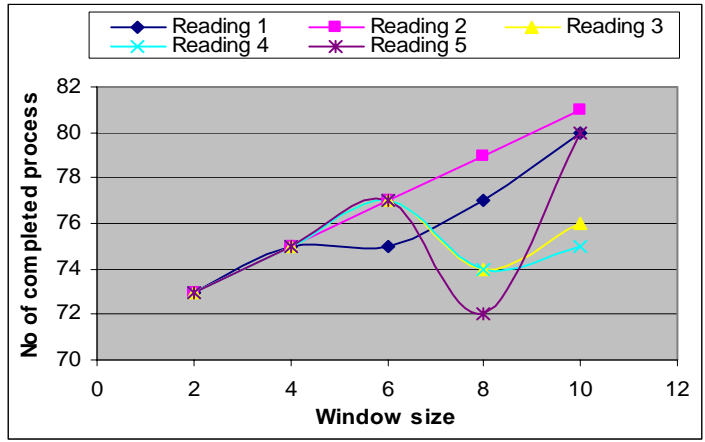

Figure. 2(b).   Performance in implementing Myopic algorithm for W=1.0.

It is noted in Figure 2(a) & 2(b) that the number of average completed processes gets higher with the larger window size with significant level of oscillations due to random nature of the processes. Figure 2(b) depicts similar level and trend of performance as Figure 2(a).

*Case 3: For 1000 processes with processing time 10 to 11 time unit and Laxity 100 time Units*

Figure 3(a) and 3(b) show the performance in implementing the Myopic algorithm for 1000 processes with processing time 10 to 11 time unit and the value for W is 0.5 and 1.0 respectively.

It is observed in Figure 3(a) that three out of five observations show a significant level of oscillations for Window size over 4. It is also noted that the oscillations are the result of equally likely processes and wider window. Figure 3(b) reflects the identical nature of the oscillation as shown in Figure 3(a), where the degree of oscillation is significant for higher Window size.

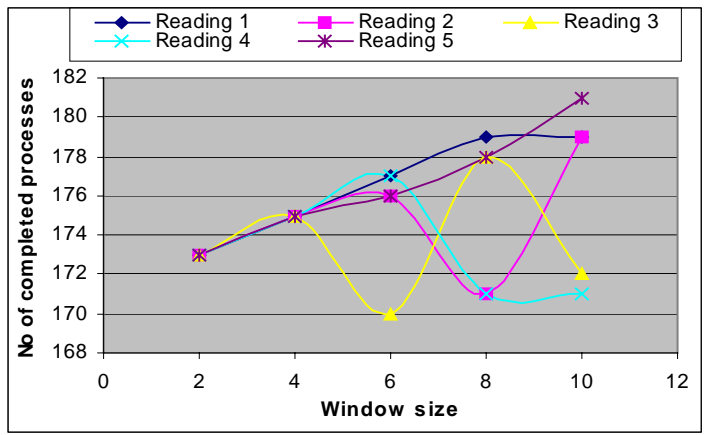

Figure. 3(a).   Performance in implementing Myopic algorithm for W=0.5.

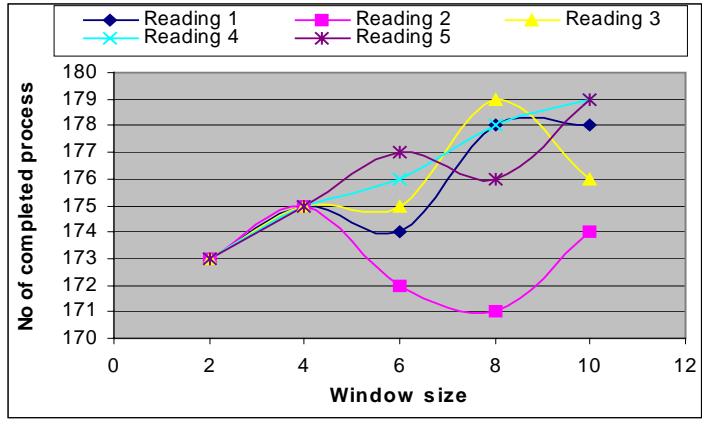

Figure. 3(b).   Performance in implementing Myopic algorithm for W=1.0.

It can be perceived from the above three cases that the algorithm achieved best performance for higher number of processes and the performance increase for larger window sizes. It implies that the Original algorithm should show the best performance. But since Original algorithm has higher complexity $O(n^2)$ than that of Myopic $O(k \times n)$, it spends more time in selecting an appropriate process that can have a bad impact on scheduling processes of short duration. This issue has been explored in the following section.

*F.  Effect of Processing Time*

This section presents the impact of Window size due to variation of processing time. To demonstrate the impact, a load of 500 processes for 5 to 6 time unit and 20 to 21 processing time are used. These are described below.

*Case 1: For 500 processes with processing time 5 to 6 time unit and laxity 100 time Units*

Figure 4(a) and 4(b) show the performance of the Myopic algorithm for 500 processes with processing time 5 to 6 time unit and the value for $W$ is 0.5 and 1.0 respectively.

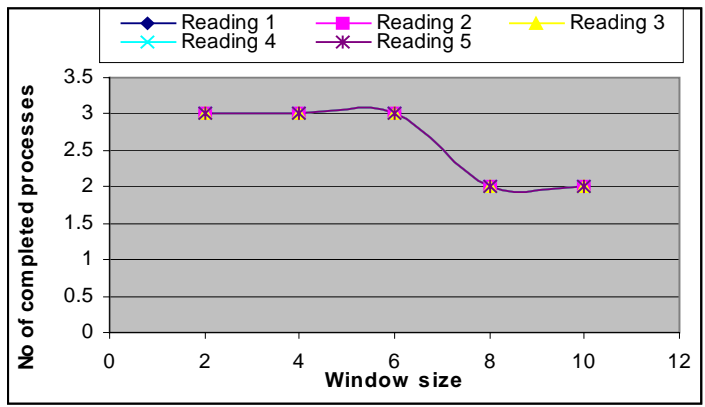

Figure. 4(a).   Performance in implementing Myopic algorithm for W=0.5.

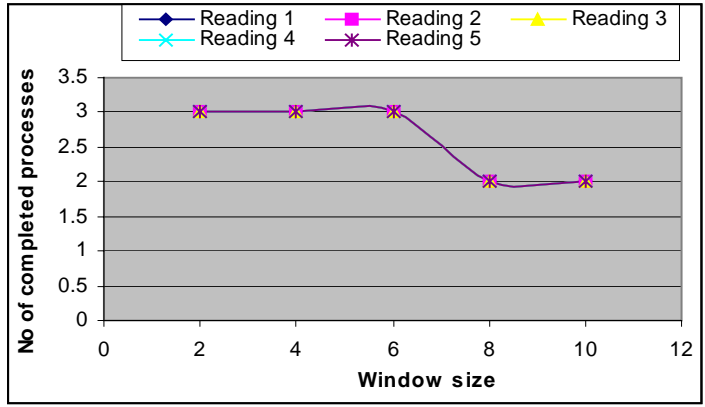

Figure. 4(b).   Performance in implementing Myopic algorithm for W=1.0.

In this case, the scheduling time dominates the processing time and more time is spent for scheduling rather than executing the processes. So, the number of completed processes drops to 3 in both cases W=0.5 and W=1.0. Since $T_P$ is small compared to scheduling time, the effect of $T_{EST}$ is insignificant in choosing a process. Figure 4(a) and 4(b) depict the same performance in terms of number of completed processes.

*Case 2: For 500 processes with processing time 20 to 21 time unit and laxity 100 time Units*

Figure 5(a) and 5(b) depict the performance in implementing the Myopic algorithm for 500 processes with processing time 20 to 21 time unit and the value for $W$ is 0.5 and 1.0 respectively.

Figure 5(a) shows that for longer processes duration, the performance (number of completed processes) is independent of window size. Similar level of performance is also observed in Figure 5(b), for larger value of $W$. Thus it is demonstrated that the value of $W$ does not have any impact on performance for processes of higher processing time.

Finally, it is observed from the above two cases that the algorithms achieved better performance for higher processing time without any impact of window size of the processes.

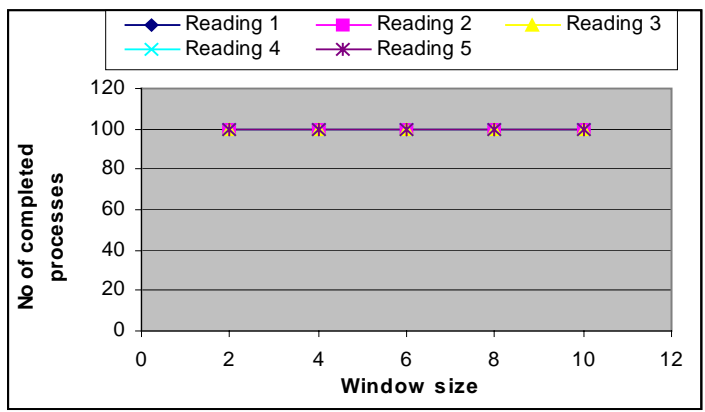

Figure. 5(a).   Performance in implementing Myopic algorithm for W=0.5.

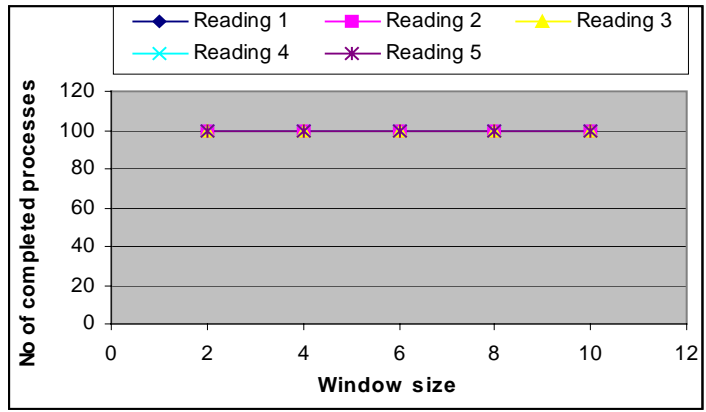

Figure. 5(b).    Performance in implementing Myopic algorithm for W=1.0.

IV.   CONCLUSION

This paper has presented the impact of the performance in implementing the Myopic algorithm for different Window sizes. A set of experiments have been performed to demonstrate the performance issues of the algorithm. It is noted that the window size plays a vital role on the performance of the Myopic algorithm, in particular, for processes of lower processing time. For relatively large number of processes and lower processing time, the algorithm achieved better performance and this increases further for larger Window size. It is also noted that the algorithm achieved better performance for the higher processing time, without any impact of the window size.

Finally, it is worth noting that the processing time and window size has a significant impact on the performance in implementing the Myopic algorithm even in a uniprocessor computing domain.